\documentclass{article}
\usepackage{amssymb}
\usepackage{fullpage}
\usepackage{todonotes}
\usepackage{graphicx}
\newcommand{\QED}{\hfill \hbox{\hskip 4pt
             \vrule width 5pt height 6pt depth 1.5pt\hskip 2pt}}

 \newtheorem{theorem}{Theorem}
 
 \newtheorem{lemma}{Lemma}
 \newtheorem{observation}{Observation}
 \newtheorem{claim}{Claim}
\newtheorem{example}{Example}
\newtheorem{definition}{Definition}

\newtheorem{fact}{Fact}

\begin{document}
\title{The k-mappability problem revisited
\thanks{This work was partially supported by ISF grant 1475/18 and BSF 
grant 2018141.}
\thanks{This work is part of the second author's Ph.\,D.\, dissertation.}}

\author{
\begin{tabular}{ccc}
Amihood Amir\thanks{ Department of Computer Science, Bar-Ilan
University, Ramat-Gan 52900, Israel, +972 3 531-8770; {\tt
amir@cs.biu.ac.il}. Partly supported by ISF grant 1475/18 and BSF 
grant 2018141} & 
Itai Boneh\thanks{Department of Computer Science, Bar-Ilan U., 52900
  Ramat-Gan, Israel,  {\tt itai.bone@biu.live.ac.il}. Partly supported by 
ISF grant 1475/18.} &
Eitan Kondratovsky\thanks{Department of
Computer Science, Bar-Ilan U., 52900 Ramat-Gan, Israel,
(972-3)531-7874, {\tt e2kondra@uwaterloo.ca}. Partly supported by ISF
grant 1475/18}  
\\
{\small Bar-Ilan University}& {\small Bar-Ilan University}& {\small
  Bar-Ilan University}\\
{\small and} & & {\small and}\\
{\small Georgia Tech} & &{\small Waterloo University}\\
\end{tabular}
}

\date{}

\maketitle

%\author{Eitan Kondratovsky}{Cheriton School of Computer Science, Waterloo University, Waterloo, Canada}{e2kondra@uwaterloo.ca}{https://u.cs.biu.ac.il/~kondrae/}{Partly
%  supported by ISF grant 1475/18}

%\ccsdesc[100]{Pattern matching, Sorting and searching}
%\keywords{Pattern Matching, Hamming Distance, Suffix Tree, Suffix Array}

%\funding{(Optional) general funding statement \dots}%optional, to capture a funding statement, which applies to all authors. Please enter author specific funding statements as fifth argument of the \author macro.

%\nolinenumbers %uncomment to disable line numbering

%\hideLIPIcs  %uncomment to remove references to LIPIcs series (logo, DOI, ...), e.g. when preparing a pre-final version to be uploaded to arXiv or another public repository

%Editor-only macros:: begin (do not touch as author)%%%%%%%%%%%%%%%%%%%%%%%%%%%%%%%%%%
%\EventEditors{Pawe{\l} Gawrychowski and Tatiana Starikovskaya}
%\EventNoEds{2}
%\EventLongTitle{32nd Annual Symposium on Combinatorial Pattern Matching (CPM 2021)}
%\EventShortTitle{CPM 2021}
%\EventAcronym{CPM}
%\EventYear{2021}
%\EventDate{July 5--7, 2021}
%\EventLocation{Wroc{\l}aw, Poland}
%\EventLogo{}
%\SeriesVolume{191}
%\ArticleNo{17}
%%%%%%%%%%%%%%%%%%%%%%%%%%%%%%%%%%%%%%%%%%%%%%%%%%%%%%

\maketitle
\begin{abstract}
The $k$-mappability problem has two integers parameters $m$ and $k$. For every subword of size $m$ in a text $S$, we wish to report the number of indices in $S$ in which the word occurs with at most $k$ mismatches. 

The problem was lately tackled by Alzamel et al. ~\cite{DBLP:conf/spire/AlzamelCIKPRS18}. For a text with constant alphabet $\Sigma$ and $k \in O(1)$, they present an algorithm with linear space and $O(n\log^{k+1}n)$ time. For the case in which $k = 1$ and a constant size alphabet, a faster algorithm with linear space and $O(n\log(n)\log\log(n))$ time was presented in ~\cite{DBLP:journals/tcs/AlzamelCIPRS20}. 

In this work, we enhance the techniques of ~\cite{DBLP:journals/tcs/AlzamelCIPRS20} to obtain an algorithm with linear space and  $O(n \log(n))$ time for $k = 1$. Our algorithm removes the constraint of the alphabet being of constant size. We also present linear algorithms for the case of $k=1$, $|\Sigma|\in O(1)$ and $m=\Omega(\sqrt{n})$.
\end{abstract}
\section{Introduction}
Many real world applications need to identify events that
repeat very often. Examples of such applications are road traffic peaks~\cite{fha:conj}, load peaks on web servers~\cite{web:02}, monitoring events in computer networks~\cite{bhivw:07}, life event histories~\cite{rbs:13} and many others. Finding such events often leads to useful insights by shedding light on the structure of the data, and giving a basis to predicting future events and behavior. Moreover, in some applications frequent events can point out a problem. In a computer network, for example, repeating error messages can indicate a misconfiguration, or even a security intrusion such as a port scan \cite{MH:ICDE01}.

In Stringology, the problem of counting the occurrences of every subword of length $m$ that appears in text $S$ is a well-known exercise in the power of suffix trees~\cite{W-73} or suffix arrays~\cite{MM-90,KS-03}. However, in reality one seldom finds {\em exact} repetitions of a substring. The situation becomes more complex when we seek the most frequent subword that {\em approximately} occurs in the string. 

Let $S[1\ldots n]$ be a text and $k$ and $m$ two integers. The {\em $k$-mappability problem} is defined as follows:
\begin{definition}
For every index $i \in [1\ldots n - m + 1]$, report the number of indices $j$ such that $HD(S[i\ldots i + m - 1], S[j \ldots j + m - 1]) \le k$. With $HD(X,Y)$ denoting the Hamming distance between $X$ and $Y$.
\end{definition}

The $k$-mappability problem was lately tackled by Alzamel et al. ~\cite{DBLP:conf/spire/AlzamelCIKPRS18}. For a text with constant alphabet and $k \in O(1)$, they present an algorithm with linear space and $O(n\log^{k+1}n)$ time. Additionally, they present a quadratic algorithm for reporting the $k$-mappability for a fixed value of $k$ and every $m\in [k\ldots n]$ or a fixed value of $m$ and every $k\in [0\ldots m]$. Finally, they show that the $k$-mappability problem can not be solved in truly subquadratic time unless the Strong Exponentional Time Hypothesis is false. For the case in which $k = 1$ and a constant size alphabet, a faster algorithm with linear space and $O(n\log(n)\log\log(n))$ time was presented in ~\cite{DBLP:journals/tcs/AlzamelCIPRS20}. ~\cite{DBLP:journals/tcs/AlzamelCIPRS20} also presented an algorithm with average case linear time for $k=1$, and provided some experimental results.

\textbf{Our results:}
\begin{enumerate}
    \item By enhancing the techniques of ~\cite{DBLP:journals/tcs/AlzamelCIPRS20}, we construct an algorithm for $k$-mappability with linear space and $O(n\log n)$ time for $k=1$ and infinite integer alphabet. This is an improvement over the $O(n\log^{k+1}n)$ time achieved by ~\cite{DBLP:conf/spire/AlzamelCIKPRS18} for $k \in O(1)$. It also improves the faster $O(n\log(n)\log\log(n))$ time for $k=1$ achieved by ~\cite{DBLP:journals/tcs/AlzamelCIPRS20}. In the settings in which infinite integer alphabet is allowed, our algorithm is optimal.
    \item We present a linear time algorithm for $k$-mappability in the case in which $k=1$, the alphabet size is constant and $m\in \Omega (\sqrt{n})$.
    %\item We present a linear time algorithm for $k$-mappability in the case in which $k=1$, the alphabet size is constant and the input string $S$ has a $t$-Anticover for a sufficiently small value $t$. Anticovers were lately presented by ~\cite{acdgikw:20} as an abstract measure indicating the lack of repetitively in a string. We provide the first evidence for the usefulness of Anticovers in constructing efficient algorithms.
\end{enumerate}

The paper is organized as follows. In Section~\ref{s:pre} we define the basic notions used. Section~\ref{s:constk} presents a linear space $O(n\log n)$ time algorithm for $1$-mappability. In Section ~\ref{s:lspace} we present a linear algorithm for $1$-mappability with constant sized alphabet and $m\in \Omega(\sqrt{n})$.

\section{Preliminaries}\label{s:pre}

Let $\Sigma$ be an alphabet. A {\em string} $S$ over $\Sigma$ is a finite
sequence of characters from $\Sigma$.
By $S[i]$, for $1\leq i\leq |S|$, we denote the $i^{th}$
character of $S$.
The {\em empty string} is denoted by $\epsilon$.
By $S[i \ldots j]$ we denote the string $S[i]\ldots S[j]$ called a {\em
  substring}, or {\em factor}, of $S$ (if $i > j$, then the substring is the empty string). 
A substring is called a {\em prefix} if $i=1$ and a {\em suffix} if $j=|S|$.
The prefix of length $j$ is denoted by $S[\ldots j]$,
while by $S[i\ldots ]$ we denote the suffix which starts from index $i$ in $S$.
We say that a string $S$ of length $n$ has a \textit{period} $p$, for some $1 \le p \le \frac{n}{2}$ if $S[i] = S[i+p]$ for every $i \in [1 \ldots n-p]$. {\em The period} of $S$, denoted as $per(S)$, is the smallest $p$ that is a period of $S$. 
We say that a substring of $S$, denoted as $A = S[a \ldots b]$ is a \textit{run} with period $p$ if its period is $p$, but $S[a-1] \neq S[a - 1 + p]$ and $S[b+1] \neq S[b+1 - p]$. This means that no substring containing $A$ has a period $p$. 
The {\em Hamming distance} of two $n$-length strings, $S_1$, and $S_2$, denoted as $HD(S_1,S_2)$, is the number of indices in which they differ. We say that an $m$-length word $w$ has a $k$-ham occurrence in location $i$ of string $S$ if $HD(w,S[i \ldots i + m - 1]) \le k$.

The longest common prefix (suffix) of two indexes $i,j \in [1\ldots n]$ is the maximal integer $\ell$ such that $S[i \ldots i +\ell - 1] = S[j \ldots j + \ell -1]$ ($S[i - \ell + 1 \ldots i] = S[j - \ell + 1 \ldots j]$). We denote the $LCP(i,j)=\ell$ ($LCS(i,j)=\ell$). LCP and LCS are collectively referred to as longest common extensions (LCE).

The {\em suffix tree}~\cite{W-73} is a useful string data structure. 
\begin{definition}\label{d:trie}
Let $S_1,\ldots,S_k$ be strings over alphabet $\Sigma$ and let
$\$\not\in \Sigma$. 

An {\em uncompacted trie of strings} $S_1,\ldots,S_k$ is an edge-labeled
tree with $k$ leaves. Every path from the root to a leaf corresponds
to a string $S_i$ with a $\$$ symbol appended to its end. The edges on this path are labeled by the symbols of
$S_i$. Strings with a common prefix start at the root and follow the
same path of the prefix, the paths split where the strings differ.

A {\em compacted trie} is the uncompacted
trie with every chain of edges connected by degree-2 nodes
contracted to a single edge whose label is the concatenation of the
symbols on the edges of the chain.

Let $S=S[1],\ldots,S[n]$ be a string over alphabet $\Sigma$. Let
$\{S_1,\ldots,S_n\}$ be the set of suffixes of $S$, where
$S_i=S[i\ldots ],\ \ i=1,\ldots,n$. A {\em suffix tree}
of $S$ is the compacted trie of the suffixes $S_1,\ldots,S_n$.
\end{definition}

For every node $u$, we call the concatenation of the labels on the path from the root to $u$ the \textit{locus} of $u$ denoted as $\mathcal{L}(u)$. For an edge $e$ in the compact trie, we use the same notation $\mathcal{L}(e)$ to denote the label (or the locus) of $e$.  Finally, for a downwards path $P$ in the compact trie, the locus $\mathcal{L}(P)$ is the concatenation of the loci of the edges in $P$. In a compact trie, an edge $e$ can have label s.t. $|\mathcal{L}(e)| > 1$. We refer to the symbol $\mathcal{L}(e)[1]$ as the symbol of $e$.

\begin{theorem}\label{t:W-3} {\em [Weiner~\cite{W-73}]}
For finite alphabet $\Sigma$, the suffix tree of a length-$n$ string
can be constructed in time $O(n)$. For general alphabets it can be
constructed in time $O(n\log \sigma)$, where $\sigma = {\rm
  min}(|\Sigma|, n)$.
\end{theorem}

The suffix tree can be preprocessed in $O(n)$ \cite{bf:04} to be used as a data structure for $LCE$ queries with $O(1)$ query time.

We assume that every node $u$ in the suffix tree contains some auxiliary information about $\mathcal{L}(u)$, that is the number of occurrences of $\mathcal{L}(u)$ in the text $S$ and a pointer to the list of indices in which $\mathcal{L}(u)$ occurs. This information can be evaluated for all the nodes of a given suffix tree $ST$ in $O(|ST|)$ time and require an additional $O(|ST|)$ space.

Over finite alphabet $\Sigma$, the adjacency list of a node $u\in ST$ is represented as an array $A_u[1\ldots |\Sigma|]$ with the edge with symbol $\sigma \in \Sigma$ in $A_u[\sigma]$ (or an emptiness indicator if there is no edge with that symbol).

Over infinite alphabet, the adjacency list of $u\in ST$ is represented as a balanced search tree storing the edges emerging from $u$ in a sorted order of their symbols. In our algorithm, we assume that the representation of the adjacency list allows linear time DFS iteration on the subtree rooted in a node $u \in ST$. This is indeed the case for most balanced trees. 

\begin{definition}\label{d:sarray}
The {\em suffix array} of a string $S$, denoted as $SA(S)$, is an integer
array of size $n+1$ storing the starting positions of all
(lexicographically) sorted non-empty suffixes of $S$, i.e.~for all  
$1 < r \leq n+1$ we have $S[SA(S)[r-1] .. n] < S[SA(S)[r] ..
  n]$. Note that the empty suffix is explicitly added to the array. 
\end{definition}  

The suffix array of $S$ corresponds to a pre-order traversal of all the
leaves of the suffix tree of $S$. Various algorithms exist for efficient time and space construction of the suffix array\cite{s:dcc:98,KS-03,FGGV06}. In particular, the suffix array over a fixed finite alphabet can be constructed in linear time.

\section{$O(n\log(n))$ time and $O(n)$ space algorithm for $k = 1$ }\label{s:constk}

\subsection{An overview of the $O(n\log(n)\log\log(n))$ algorithm for $k=1$}
We start with an overview of the ideas for the $O(n\log(n)\log\log(n))$ algorithm of  ~\cite{DBLP:journals/tcs/AlzamelCIPRS20}. They present an algorithm for counting the number of occurrences with \textbf{exactly} one mismatch, for every word of size $m$. Since there is a textbook algorithm for counting the number of exact occurrences of every word, this is sufficient for solving the $1$-mappability problem.

They start by evaluating the suffix tree $T$ of $S$ and trimming the tree at word length $m$. That is, every node $v$ with $|\mathcal{L}(v)| > m$ is removed. Implicit nodes with $|\mathcal{L}|=m$ are made explicit leaves in the trimming process. They proceed to evaluate the heavy paths decomposition of $T$.
\begin{definition}[Heavy Path Decomposition]
Let $T$ be a rooted tree. For every non-leaf vertex $u$, the edge $(u,v)$ is \textit{heavy} if $|I_u| < 2|I_v|$ with $I_x$ denoting the set of leaves in the subtree rooted in the vertex $x$. An edge that is not heavy is called a \textit{light} edge. The heavy path of a vertex $v$ is the maximal path of heavy edges going through $v$ (it may contain 0 edges). For every heavy path $P$, a vertex $u \in P$, and a light edge $(u,v)$ emerging from $u$, we call $T(v)$ a \textit{sidetree} of $P$ (emerging from $u$). 
\end{definition}
% Comment by Eitan:
% \todo[inline]{Few small notes on the heavy path definition:
% \begin{enumerate}
%     \item Heavy path def. (Definition 1 from 1-mappability paper) is different. Questions: Should we change our definition or note that the algorithm also works with this defition
%     \item $T(x)$ is confusing with $T$ (the text). We may use $|I_u|$ as used in Panos's paper about 1-mappability.
% \end{enumerate}}

It is easy to observe that every root-to-leaf path in $T$ consists of at most $\log(n)$ heavy paths and $\log(n)$ light edges. The following observation is the key for the complexity achieved by ~\cite{DBLP:journals/tcs/AlzamelCIPRS20}:

\begin{observation}\label{obs:ham1split}
For every $w= S[i\ldots i + m - 1]$, every 1-ham occurrence of $w$ $w'=w[1 \ldots x-1] \sigma w[x+1 \ldots m]$ with a mismatch in index $x$ corresponds to a node $u$ in $T$ with $\mathcal{L}(u) = w[1\ldots x-1]$. $u$ must have two edges $e_1,e_2$ s.t. there is a downwards path starting with $e_1$ (resp. $e_2$) and ending in a leaf with path label $w[x \ldots m]$ (resp. $\sigma w[x+1 \ldots n]$).
\end{observation}

Consider the following procedure:
For every node $u\in T$ with path label $w$, let the heavy edge emerging from $u$ be $e_h$ with label $d$. Inspect every light edge $e=(u,v)$ with label $c$ emerging from $u$. For every leaf $z\in T(v)$ with label $L(z) = w \cdot c \cdot w_z$ and for every $c' \neq c \in \Sigma$, find the leaf $z'$ with label $\mathcal{L}(z') = w \cdot c' \cdot w_z$, if it exists. If it does, add the number of occurrences of $\mathcal{L}(z')$ to a counter associated with $z$. For the leaf $z_d$ with $\mathcal{L}(z_d) = w \cdot d \cdot w_z$, also increment a counter associated with $z_d$ by the amount of occurrences of $\mathcal{L}(z)$.

It is straightforward from Observation \ref{obs:ham1split} that for every index $i$, every $1$-ham occurrence is counted by the above procedure. As for complexity - every leaf $z$ is iterated once per light edge in the path from the root to $z$. A single iteration on a leaf $z$ consists of a constant number of counter increments and a single query for finding $z'$ with $L(z') = w \cdot c' \cdot w_z$ per symbol $c' \in \Sigma$. Since $|\Sigma| = O(1)$, the bottleneck of the iteration is finding $z'$. The following is proven in ~\cite{DBLP:journals/tcs/AlzamelCIPRS20}:

\begin{theorem}\label{t:conque}
A text $S[1\ldots n]$ can be preprocessed in time $O(n\log\log n)$ and linear space to allow the following query in $O(\log\log n)$ time: 

Given a node $u$ in the suffix tree of $S$ with $L(u) = w_1 \cdot c \cdot w_2$ ($w_1,w_2 \in \Sigma^*$ and $c \in \Sigma$) and a symbol $c \neq c' \in \Sigma$, find the node $u'$ with $L(u') = w_1 \cdot c' \cdot w_2$ if it exists.
\end{theorem}

We call the queries described in Theorem \ref{t:conque} \textit{concatenation} queries.

With Theorem \ref{t:conque} the final complexity is clear - every leaf is iterated $O(\log n)$ times and the iteration costs $O(\log\log n)$ after an $O(n\log\log n)$ preprocessing time. The overall time complexity is $O(n\log(n)\log\log(n))$

\subsection{Linear space $O(n\log n)$ algorithm for $k=1$}\label{s:1km}
\textbf{Intuition:} Our algorithm is based on the ideas of ~\cite{DBLP:journals/tcs/AlzamelCIPRS20}. For every light edge $(u,v)$ we iterate every leaf $z\in T(v)$ and wish to find the vertices corresponding to a 1-ham occurrence of $\mathcal{L}(z)$ with a mismatch in index $|\mathcal{L}(u)| + 1$. Instead of using concatenation queries, we construct a lexicographically sorted array of the words $W$ we need to find. Given the sorted array of words, finding the vertices corresponding to these words in the suffix tree can be done in $O(|W|)$. If we manage to construct this sorted array in $O(|W|)$, the amortized time for inspecting a leaf is constant (rather than $O(\log\log n)$).

\textbf{Terminology}. Let $P=(u_1,u_2, \ldots u_x)$ be a heavy path in the heavy path decomposition of the suffix tree $ST$ of $S$. Let $\mathcal{L}(u_i) = w_i$ and let $e_i = (u_i, u_{i+1})$ be the $i$'th heavy edge in $P$ with symbol $d_i$. Let $(u_i,v)$ be a light edge emerging from $u_i$ with symbol $c$ and let $z\in T(v)$ be a leaf. It holds that $\mathcal{L}(z) = w_i \cdot c \cdot s_z$ for some suffix $s_z \in \Sigma^*$.

\begin{definition}
The node $z'\in ST$ is a \textit{$P$-light occurrence} of $z$ if $L(z') = w_i \cdot c' \cdot s_z$ for some $c' \in \Sigma \setminus \{c,d_i \}$. We call the word $hw(z) = w_i \cdot d_i \cdot s_z$ the \textit{$P$-heavy word} of $z$. The node $z'\in ST$ is a \textit{$P$-heavy occurrence} of $z$ if $L(z') = hw(z)$.
\end{definition}

Note that the above definitions are with respect to a heavy path $P$. $z$ may be a leaf in the sidetrees of multiple heavy paths. In every such path, the $P$-light occurrences, $P$-heavy occurrence and the $P$-heavy word of $z$ are different. Also note that the $P$-heavy word and the $P$-heavy occurrence are undefined for leaves in the sidetrees emerging from $u_x$, as the last heavy edge in $P$ is $e_{x-1}$.

In our algorithm, we count the $P$-heavy occurrences and the $P$-light occurrences of every node $z$ in a sidetree of $P$ independently. For every heavy occurrence $z'$, we also count the occurrences of $w(z)$ as $1$-ham occurrences of $w(z')$. We do this for every heavy path $P$. Surely, this process counts all the $1$-ham occurrences.

We start by showing how to efficiently count the $P$-light occurrences.

\begin{observation}\label{obs:lighoc}
For every vertex $z$ with $\mathcal{L}(z) = w$ in a sidetree $T(v)$ emerging from $u_i$, all the $P$-light occurrences of $z$ are also leaves in (different) sidetrees emerging from $u_i$. Furthermore, a leaf $z'$ with $\mathcal{L}(z') = w'$ in a sidetree $T(v') \neq T(v)$ emerging from $u_i$ is a light occurrence of $z$ iff $w[|w_i + 2| \ldots m] = w'[|w_i+2|\ldots m]$.
\end{observation}

Observation \ref{obs:lighoc} is directly derived from the definition of a light occurrence. For every $u_i\in P$, we wish to construct a sorted array consisting of the suffixes starting in index $|w_i| + 2$ of the labels of the leaves of the sidetrees emerging from $u_i$.

We present the following routine: 

{\bf Suffix Sorting:}\\
\fbox{\begin{minipage}{12cm}
 As a preprocess procedure, construct the suffix array $SA$ of $S$. 
\\\\
{\bf Initialize} an array $A$ of size $n$ consisting of empty lists
\\\\
\textbf{ Alignment step:} Iterate the leaves in the sidetrees of $u_i$. For every leaf $z$, extract $j_z$ - a starting index of $L(z)$. We add $z$ to $A[j_z + |w_i| + 2]$.
\\\\
{\bf Insertion step:} Initialize an empty list $L$. Iterate $SA$ from left to right. When iterating $SA[j]$, add all the nodes in $A[SA[j]]$ to the end of $L$.
\end{minipage}}

\begin{claim}\label{c:sorLight}
After running Suffix Sorting, $L$ is sorted by the lexicographic order of the suffixes starting in index $|w_i| +2$ of $z$. The running time is $O(n + |SE|)$ with $SE$ being the set of sorted elements.
\end{claim}
\textbf{Proof:} leaf $z$ with $\mathcal{L}(z)=w_z$ occurring in index $j_z$ is inserted to $L$ before the leaf $y$ with $L(y) = w_y$ occurring in index $j_y$ only if the suffix of $S$ starting in $j_z + |w_i| + 2$ is lexicographically smaller or equal to the suffix of $S$ starting in $j_y + |w_i| + 2$. Therefore, it can not be the case that $w_z[|w_i| + 2 \ldots m] >_L w_y[|w_i| + 2 \ldots m]$.
\\\\
As for complexity- the alignment step takes $O(|SE|)$ time as it executes a constant amount of list insertions and basic arithmetic operations for every leaf. The insertion step takes $O(|SE| + n)$ time as it iterates over the entire suffix array. The sum of the sizes of the lists in $A$ is identical to the amount of iterated leaves in the alignment step. We assume that the suffix array was evaluated prior to the run of Suffix Sorting. Therefore, we exclude the complexity of computing the suffix array from our running time.   \QED

This is not exactly what we want. If we  execute Suffix Sorting for every node, the $n$ factor will dominate the complexity and the overall time will be quadratic. To avoid that, we present the following algorithm for sorting a batch of sidetrees. 

{\bf Batched Suffix Sorting:}\\
\fbox{\begin{minipage}{12cm}
{\sf As a preprocess procedure, construct the suffix array $SA$ of $S$. }. 
\\\\
{\sf Input:} A batch of vertices $v_1,v_2 \ldots v_b$ 
\\\\
{\sf Initialize} an array $A$ of size $n$ consisting of empty lists.
\\\\
{\sf Batched Alignment step:}
\\\\
{\sf For every} $i \in [1\ldots b]$:
\begin{enumerate}
    \item {\sf Initialize} an empty list $L_i$
    \item Iterate the leaves in the sidetrees of $v_i$ with $L(v_i) = w_i$. For every leaf $z$, extract $j_z$ - a starting index of $\mathcal{L}(z)$. Add the pair $(z,L_i)$ to $A[j_z + |w_i| + 2]$. 
\end{enumerate}
{\bf Batched Insertion step:}  Iterate $SA$ from left to right. When iterating $SA[j]$, for every  $(z,L) \in A[SA[j]]$, add $z$ to the end of $L$.
\end{minipage}}

The same arguments as in the proof of claim \ref{c:sorLight} can be made to prove the following:
\begin{claim}
After running Batched Suffix Sort, every list $L_i$ has the leaves in the sidetrees of $v_i$ sorted by the lexicographic order of the suffixes starting in index $|w_i| +2$ of $z$. The running time is $O(n +|SE|)$ with $SE$ being the set of sorted elements in the batch.
\end{claim}

To sort the sidetrees in amortized linear time, we set a counter $se=0$ for the amount of leaves in the sidetrees that need to be sorted and an empty list $Sort$. We iterate the vertices in $ST$. For every vertex $u$, we count the number of leaves in the sidetrees of $u$, add this number to $se$ and add $u$ to $Sort$. Once $se > n$, we execute Batched Suffix Sort on $Sort$.

Since the number of leaves in the sidetrees of a vertex $u$ never exceeds $n$, it is guaranteed that $se \le 2n$ when we execute Batched Suffix Sorting. Therefore, the overall complexity is $O(n + 2n) = O(n)$. Since we only execute the batched insertion with $se \ge n$, the amortized time for placing every leaf in the sorted list is constant.

Once we have the sorted list $L=z_1,z_2 \ldots z_t$ of the leaves in the sidetrees emerging from $u_i \in P$, a simple iteration can be implemented to count the number of $P$-light occurrences for every node in $L$. We start by preprocessing $S$ for constant time lcp queries. We iterate $L$. For every consecutive pair of leaves $z_a$ and $z_{a+1}$ with $\mathcal{L}(z_a) = s_1$ occurring in $j_1$ and $\mathcal{L}(z_{a+1}) = s_2$ occurring in $j_2$, we query $l = lcp(j_1 + |w_i| + 2,j_2 + |w_i| + 2)$. If we have $l \ge  m - |w_i| - 1$, then $z_{a}$ and $z_{a+1}$ are $P$-light occurrences of each other (Observation \ref{obs:lighoc}). Once we identify a pair $z_a,z_{a+1}$ of $P$-light occurrences, we proceed in $L$ until we reach a leaf $z_{b+1}$ that is not a $P$-light occurrence. Of course, all the pairs $z_x,z_y$ with $x \neq y$ and $x,y \in [a \ldots b]$ are $P$-light occurrences of each other. We evaluate the sum $Oc$ of occurrences of $\mathcal{L}(z_x)$ for $x \in [a\ldots b]$ and increment the counter of ham-1 occurrences of $z_x$ by $Oc - Oc(z_x)$ with $Oc(z_x)$ being the number of occurrences of $\mathcal{L}(z_x)$.

It can be easily verified that the iteration is linear. For every leaf we execute a single lcp query and a constant number of basic arithmetic operations. We conclude the handling of $P$-light occurrences with the following theorem:

\begin{theorem}\label{t:fastlight}
The $1$-ham occurrences of $\mathcal{L}(z)$ that are corresponding to $P$-light occurrences of some heavy path $P$ can be computed for every leaf $z\in ST$, in $O(n\log n)$ time and linear space.
\end{theorem}

\textbf{Proof:}
For every heavy path $P$ and vertex $u_i \in P$, we compute the sorted list of the suffixes starting in $|w_i| + 2$ of the words of the leaves of the sidetrees emerging from $u_i$. We use the sorted list to find the $P$-light occurrences of the leaves in the sidetrees of $u_i$. Sorting the leaves is done using Batched Suffix Sorting with batches of size between $n$ and $2n$ and takes a constant amortized time per sorted leaf. There may be one 'remainder' batch with size $se < n$ that takes an additional $O(n)$ time to sort. Given the sorted lists, finding the $P$-light occurrences is linear in the number of leaves in the sidetrees of $u_i$. Every leaf $z$ participates in at most $\log(n)$ different sidetrees, so the overall time is $O(n\log n)$.  We also build the Suffix array as a preprocess step, which takes an additional $O(n\log n)$ time.

As for space - the only non-trivially linear part of our solution is the array $A$ used in Batched Suffix Sort. Since we never let $se$ the number of sorted elements exceed $2n$, the lists in $A$ never contain more than $2n$ elements collectively. So the size of $A$ is always linear. After executing the Batched Suffix Sorting, we iterate the sorted lists to count the $P$-light occurrences and then reuse the space occupied by these lists as they are no longer required. \QED

We are left with the task of counting the $P$-heavy occurrences of every leaf $z$. Consider a heavy path $P=u_1,u_2 \ldots u_x$. Our key sub-task for finding all the $P$-heavy occurrences of all the leaves in the sidetrees of $P$ is constructing a sorted list of the $P$-heavy words of the leaves.

Note that unlike $P$-light occurrences, $P$-heavy occurrences of a leaf $z$ of a sidetree emerging from $u_i$ can not be in a sidetree emerging from $u_i$. However, they must be leaves of a sidetree emerging from $u_j$ for some $j > i$.

The process of building the sorted list of $P$-heavy words relies on the same principles we used for the $P$-light occurrences. However, there is a further difficulty to tackle. With $P$-light occurrences that lie on the same $u_i$ - we have a guarantee that the words match until the index $|w_i|$. Therefore, it is sufficient to sort by the suffixes starting right after the mismatch in index $|w_i| +2$. With the $P$-heavy words, we may have to compare $P$- heavy words from sidetrees of different nodes $u_i$ and $u_j$ with $i < j$. In this case, there is no guarantee that the words match in the indices in $[|w_i + 1| \ldots |w_j|]$.

To handle this difficulty, we partition the leaves into classes prior to sorting them. Our partition will have the property that the $P$-heavy words of leaves in the same class have a certain common prefix that exceeds the index in which the error occurs ($c$ is replaced by $d_i$). This property will allow us to sort the $P$-heavy words in every class using Batched Suffix Sorting. 

The first step for sorting the $P$-heavy words is to partition the leaves in the sidetrees of $P=u_1,u_2 \ldots u_x$ by the $lcp$ of their $P$-heavy words with $w_x$. This is done with the following procedure: 

{\bf LCP Partition:}\\
\fbox{\begin{minipage}{12cm}

{\sf Input:} A heavy path $P = u_1,u_2 \ldots u_x$ 
\\\\
{\sf Initialize} an array $LCP[1\ldots m]$ of size $m$ of empty lists. Let $j_x$ be an index in which $w_x$ occurs.
\\\\
\textbf{Alignment Step:}
{\sf For every $i \in [1\ldots x]$:}
For every leaf $z$ in a sidetree emerging from $u_i$:
\begin{enumerate}
    \item Extract an index $j_z$ in which $\mathcal{L}(z)$ occurs in $S$. 
    \item \label{step:findlz} Find $l_z = lcp(hw(z),w_x)$ by computing $l_z = min(|w_i| + 1 + lcp(j_z + |w_i| + 2,j_x + |w_i| + 2),|w_x|)$.
    \item Compare between the symbols in index $l_z+1$ in $hw(z)$ and in $w_x$ in order determine the lexicographical order $o_z \in \{ < ,> ,=\}$ between $hw(z)$ and $w_x$ (For example, $o_z = <$ if $hw(z) <_L w_x$). If $l_z = |w_x|$, $o_z$ is set to $'='$. 
    \item Add the tuple $(z,o_z)$ to $LCP[l_z]$.
    \end{enumerate}

\textbf{Insertion Step:}
\\\\
{\sf For every $l \in [1\ldots |w_x| -  1]$ (in increasing order):}  
\begin{enumerate}
    \item If the list $L = LCP[l]$  is empty - do nothing.
    \item Otherwise, create 2 lists $L_l^>$ and $L_l^<$.
    \item For every tuple $(z,o_z)$, add $z$ to $L_l^{o_z}$.
\end{enumerate}

{\sf If} $L=LCP[|w_x|]$ is not empty, construct a new list $L_{|w_x|}$ and add $z$ to $L_{|w_x|}$ for every pair $(z,=) \in L$.
\end{minipage}}

Note that $l_z \ge |w_i| + 1$ since $hw(z)[1 \ldots |w_i|] = w_i= w_x[1\ldots |w_i|]$ and $hw(z)[|w_i + 1|] = d_i$. With that observation, it is clear that the formula for finding $l_z$ in Step \ref{step:findlz} works. 

We make the following observation:

\begin{observation}\label{obs:lcplists}
For every list $L_l^>$ (or $L_l^<$ or $L_{|w_x|}$), every vertex $z\in L_l^>$ has $lcp(hw(z),w_x) = l$ and $hw(z)[l+1 \ldots m] = \mathcal{L}(z)[l+1 \ldots m]$. The running time of LCP partition is $O(m + |SE|)$ with $SE$ being the set of leaves in sidetrees of $P$.
\end{observation}
\textbf{Proof:} The $lcp$ property is derived directly from the construction of $LCP[1 \ldots m]$. As for complexity, every leaf is processed with a single $lcp$ query and a constant amount of basic operations. The iteration and construction of $LCP$ takes an additional $O(m)$ \QED

It follows from Observation \ref{obs:lcplists} that the lexicographical order between the $P$-heavy words of the vertices in $L_l^>$ (or $L_l^<$ or $L_{w_x}$) are determined by the suffixes starting in index $l+1$ of $L(z)$. Therefore, sorting $L_l^>$ by the lexicographical order of the heavy words can be done using the algorithm Batched Suffix Sorting.

\begin{theorem}\label{t:fastphwt}
The lexicographically sorted list of $P$-heavy words of a heavy path $P$ can be evaluated in $O(|SL|)$ amortized time with $SL$ the set of leaves in the sidetrees of $P$
\end{theorem}
\textbf{Proof:} We want to use LCP Partition on a batch of heavy paths. Transforming LCP Partition to a batched algorithm can be done with the same technique that was used to generate Batched Suffix Sorting from Suffix Sorting. 

As in the Batched Suffix Sorting algorithm, we execute the alignment step of LCP Partition for possibly multiple heavy paths $P$ until the collective amount of leaves considered is between $n$ and $2n$. Once this amount is met, we construct the lists $L^>_l$, $L^<_l$ and $L_{|w_x|}$ for all the paths in the batch by applying the insertion step. We then sort the lists by the lexicographic order of $hw(z)$ with Batched Suffix Sorting. The overall time is $O(n + m) = O(n)$. 

We are left with the task of merging the sorted lists $L_l^>$,$L_l^<$ and $L_{|w_x|}$ into a single sorted list $L$ containing all the $P$-heavy words. This is done by applying the following observation:

\begin{observation} \label{obs:concsort}
Let $l_1,l_2, \ldots l_c$ be the set of indices for which either $L_{l_i}^<$ or $L_{l_i}^>$ is constructed for the path $P$ by LCP Partition. The sorted list $L$ of the $P$-heavy words is of the form $L= L_{l_1}^< , L_{l_2}^<, \ldots L_{l_c}^<,L_{|w_x|},L_{l_c}^>,L_{l_{c-1}}^> \ldots L_{l_1}^>$. (If $L_{l_i}$ was not constructed, it is considered as an empty list) 
\end{observation}
\textbf{Proof:} We start by showing that the lists $L_{l_i}^<$ must appear in increasing order of $l_i$ in $L$. Let $a,b \in \{ l_1,l_2 \ldots l_c\}$ be two indices for which $a<b$. Let $w_a \in L_a^<$ and $w_b \in w_b^<$. Since the LCP of $w_b$ and $w_x$ is $b \ge a + 1$, we have $w_b[a+1] = w_x[a+1]$. Since the LCP of $w_a$ and $w_x$ is $a$ and $w_a <_L w_x$, we have $lcp(w_a,w_b) = a$ and $w_a[a+1] <_L w_x[a+1] = w_b[a+1]$ and therefore $w_a <_L w_b$. Similar arguments can be made to prove that $w_b <_L w_a$ for every $w_a \in L_{l_i}^>$ and $w_b \in L_b^>$. It is straight forward from the construction of the lists $L_{l_i}^<$ and $L_{l_i}^>$ that for every $l_i$ and every $W \in L_{l_i}^>$ , $w \in L_{l_i}^<$ and $w' \in L_{|w_x|}$ we have $w <_L w' <_L W$. \QED

With Observation \ref{obs:concsort}, the construction of the sorted $P$-heavy words list is completed. Observe that LCP Partition naturally generates $L_l^>$ and $L_l^<$ in increasing order of $l$. Therefore, the concatenation of the lists in the order dictated by Observation \ref{obs:concsort} does not require any further sorting and can be executed in linear time, and the proof of Theorem \ref{t:fastphwt} is completed.\QED

Given the sorted list $L_P[1 \ldots h]$ of the $P$-heavy words, we are interested in finding the node $z'$ with $L(z') = hw(z)$ for every word $hw(z) \in L_P$. We can do this in linear time as follows: First, observe that every $P$-heavy word has the prefix $w_1$. So $z'$ ,if it exists, must be a descendant of $u_1$ and therefore is a leaf in a sidetree of $P$. Let $L[1\ldots l]$ be the sorted list of occurrences of $w_1$ stored in $u_1$. These are actually all the leaves in the sidetrees of $P$. The following procedure matches every $hw(z) \in L_P$ with its corresponding $z'$:
\\\\
{\bf Count $P$-Heavy:}\\
\fbox{\begin{minipage}{12cm}

{\sf Input:} The lexicographically sorted lists $L_P[1 \ldots h]$ of $P$-heavy words and $L[1 \ldots l]$ the list of lexicographically sorted vertexes in the sidetrees of $P$
\\\\
{\sf Initialize} two indices $i = j = 1$.
\\\\
\textbf{While $i \le h$ and $j \le l$:}
\begin{enumerate}
    \item Let $hw(z) = L_P[i]$ and $\mathcal{L}(z') = L[j]$
    \item \textbf{If $hw(z)=\mathcal{L}(z')$:} 
    \begin{enumerate}
        \item Increase the counter associated with $z$ by $Oc(z')$.
        \item Increase the counter associated with $z'$ by $Oc(z)$.
        \item Increase $i$ by $1$.
\end{enumerate}
    \item \textbf{If $hw(z) <_L \mathcal{L}(z')$:} Increase $i$ by $1$.
    \item \textbf{If $hw(z) >_L \mathcal{L}(z')$:} Increase $j$ by $1$.
\end{enumerate}
\end{minipage}}

It can be easily verified that Count $P$-Heavy counts the $P$-heavy occurrence $z'$ of every leaf $z$ in a sidetree of $P$. Notice that double counting will not occur. That is due to the following: 

\begin{fact} \label{f:incij}
Let $z$ and $z'$ be two leaves in sidetrees of $P$ emerging from $u_i$ and $u_j$ respectively such that $hw(z) = \mathcal{L}(z')$. It must be the case that $i < j$.
\end{fact}
Fact \ref{f:incij} guarantees that if we count the occurrences of $z$ as $1$-ham occurrences of $z'$ and vice versa when $hw(z)$ and $\mathcal{L}(z')$ are be visited in Count $P$-Heavy, we will not count them as $1$-ham occurrences of each other again, because it can't be the case that $hw(z') = \mathcal{L}(z)$.
\\\\
The lexicographic comparisons between $hw(z)$ and $\mathcal{L}(z')$ can be executed in constant time using $lcp$ queries. To efficiently execute an $lcp$ query with a $P$-heavy word, we store the $P$-heavy word $hw(z)$ as a pair $(z,i)$ with $i$ the index in which $\mathcal{L}(z)$ is modified. With that representation, two $lcp$ queries can be used to find $a = lcp(hw(z),\mathcal{L}(z'))$ in a 'kangooroo' jump manner. If $a < m$, the following symbol can be compared to determine the lexicographic order between $hw(z)$ and $\mathcal{L}(z')$. With the constant time lexicographic comparing, it is easy to see that the complexity of Count $P$-Heavy is $O(|L_P| + |L|) = O(|SE|)$ with $SE$ being the set of leaves in the sidetrees of $P$.

Note that when the equality $hw(z) = L(z')$ is met, it is crucial to increase $i$ rather than $j$. That is due to the fact that $L_P$ may contain duplicates while $L$ does not. Alternatively, $L_P$ can be preprocessed to group duplicates together. We conclude the counting of $P$-heavy occurrences with the following:
\begin{theorem}\label{t:fastheavy}
The $P$-heavy occurrences of every leaf $z$ can be counted over all the heavy paths $P$ such that $z$ is a leaf in a sidetree of $P$ in $O(n\log n)$ time and linear space.
\end{theorem}
\textbf{Proof:} For every heavy path $P = u_1,u_2 \ldots u_x$, we use Theorem \ref{t:fastphwt} to obtain the list $L_P$ of sorted $P$-heavy words and obtain $L$ from $u_1$. We then apply Count $P$-Heavy on $L_P$ and $P$ to match every $hw(z) \in L_P$ with its $P$-heavy occurrence $z'$ if exists, and update the corresponding counters accordingly. 

The amortized time for applying Theorem \ref{t:fastphwt} for a path $P$ is $O(|SE(P)|)$ with $SE(P)$ being the set of leaves in the sidetrees of $P$. Every leaf in $ST$ is a leaf in the sidetree of at most $\log(n)$ heavy paths, so the overall complexity is $O(n\log n)$. We also construct the suffix array as a preprocess procedure, which takes an additional $O(n\log n)$ time.

As for space, the only non-trivially linear part is the array $LCP[1 \ldots m]$ used in LCP Partition. As before, we apply LCP Partition on batches of size at most $2n$, so the collective size of the lists in $LCP[1\ldots m]$ never exceeds $O(n)$. After obtaining $L_P$ for all the paths in the batch, we apply Count $P$-Heavy for every path in the batch and then reuse the space occupied by the sorted lists $L_P$ as they are no longer required. \QED

When put together, Theorem \ref{t:fastlight} and Theorem \ref{t:fastheavy} yield the main result of this section:
\begin{theorem}\label{t:1map}
The $1$-mappability problem can be solved using $O(n\log n)$-time and linear space on a text with infinite integer alphabet.
\end{theorem}

Note that for infinite integer alphabet, better time can not be achieved unless certain values of $m$ are excluded. For example: 
\begin{observation}
For a text $S$ over infinite integer alphabet and $m=2$, there is an index $i \in [1\ldots n]$ with at least $1$-ham occurrence iff the symbols of $S \cdot \sigma'$ are not distinct for some $\sigma' \notin \Sigma$.
\end{observation}
The above straight forward observation shows a trivial relation between the $k$-mappability problem and reporting whether or not all the elements of a set are distinct - which can not be done in $o(n\log n)$. It can be easily generalized for every fixed value of $m$.

\section{$O({n^2\over m^2} + n)$ and Linear Space Algorithm for $1$-Mappability with Constant sized Alphabet}\label{s:lspace}
As a warm up, we present a technique for counting the $1$-ham occurrence of a word with size $m$ in $O(\frac{n}{m})$ time. Applying this technique to every $m$-sized word yields an $O(\frac{n^2}{m})$ algorithm for $1$-mappability. We then proceed to show how to process all the words of size $m$ not one by one, but in batches of size $O(m)$. We extend the technique used in the warm up to handle a batch in $O(\frac{n}{m} + m)$ time. Since there are $O(\frac{n}{m})$ batches, this yields an $O((\frac{n}{m})^2 + n)$ time algorithm. 
\subsection{Warm up - $O(\frac{n^2}{m} + n)$}
Let $w= S[i \ldots i+m-1]$ be a subword of $S$ with length $m$.
 \begin{definition} \label{d:loccrocc}
 Let $w_1=S[j \ldots j+m-1]$ be a $1$-ham occurrence of $w$. $w_1$ is an $l$-occurrence of $w$ if $w_1[1\ldots \lceil \frac{m}{2} \rceil ] = w[1\ldots \lceil \frac{m}{2} \rceil]$. $w_1$ is an $r$ occurrence of $w$ if $w_1[\lceil \frac{m}{2} \rceil +1 \ldots m] = w[\lceil \frac{m}{2} \rceil +1 \ldots m]$. We respectively denote as $Lo(w)$ and $Ro(w)$ the sets of $l$-occurrences and $r$-occurrences of $w$ in $S$.
 \end{definition}

It is easy to see that $|Lo(w)| + |Ro(w)| - \#w$ is the number of $1$-ham occurrences of $w$, with $\#w$ denoting the number of proper occurrences of $w$ in $S$. In this section, we show how to evaluate the number of $l$-occurrences of a given word $w$ in $O(\frac{n}{k})$ time. A symmetrical approach can be applied to count the number of $r$-occurrences of $w$. $\#w$ can be evaluated for all the subwords of $S$ in $O(n)$ time using the suffix tree.

\begin{theorem}\label{t:apRep}
All the occurrences of a string $w$ of size $m$ in a text of size $n$ can be represented by a set of $O(\frac{n}{m})$ arithmetic progressions of the form $A = (s,e,d)$ such that $A = (s,e,d)$ represent a sequence of occurrences with starting indexes $\{i_x = s + d\cdot x| x\ge 0 ,i_x \le e \}$. If $w$ is periodic, every arithmetic progression $A=(s,e,d)$ has $d =per(w)$. $|A|=e-s+1$ represents the number of occurrences represented by $A$. Every arithmetic progression that has $A > 1$  corresponds to a periodic set of instances contained within a run with period $d$. This representation is called the periodic occurrences representation of $w$ and it can be obtained in $O(\frac{n}{m})$ time from the suffix tree following an $O(n)$ time preprocessing.
\end{theorem}

A proof for the above can be found in Section \ref{A:complspace}.

Given a words $w=S[i \ldots i +  m- 1]$, we use Theorem \ref{t:apRep} to obtain all the occurrences of $w_L = w[1 \ldots \lceil \frac{m}{2} \rceil]$ in periodic occurrences representation. For every occurrence of $w_L$ in this representation, we wish to check if it is a prefix of an $l$-occurrence.

We process every arithmetic progression $A = (s,e,d)$ of occurrences of $w_L$. If $A$ only represents a single occurrence of $w_L$ in index $s$, we query $l_1 = LCP(s,i)$. If $l_1 \ge m$, we have a proper instance of $w$. Otherwise, we have a mismatch. Proceed to query $l_2 = LCP(s + l_1 + 1, i + l_1 + 1)$. If it is the case that $l_2 + l_1 + 1 \ge m$, we count $s$ as an $l$-occurrence of $w$.

If $A=(s,e,d)$ represents multiple occurrences of $w_L$, then $w_L$ must have a period $d$. We exploit the periodic structure of the occurrences represented by $A$ to compute $l_2$ for all the occurrences in $A$ using constant time. The following lemma proven in Section \ref{A:complspace} is the key for doing so.

\begin{lemma} \label{o:RepetativeLCP}
Let $A=(s,e,d)$ be an arithmetic progression representing a set of indexes $s_j = s + j \cdot d$ for $j \in [0 \ldots |A|-1]$ within a run with period $d$.

Let $i \in [1\ldots n]$ be an index and let $l_p = lcp(i,s)$. Let $Ex_i$ be the maximal extension of a run with period $d$ containing $i$ to the right of $i$ (regardless of periodicity, $Ex_i\ge d$), and let $Ex_s$ be the maximal extension of the period $d$ to the right of $s$.

\begin{enumerate}
    \item If $l_p < d$:  $LCP(i,s_j) = l_p$ for every $j \in [0 \ldots |A|-2]$.
    \item Otherwise, $LCP(i,s_j) = min(Ex_i,Ex_s - j\cdot d)$  for every $j\in [0 \ldots |A|-1]$ such that $Ex_i \neq Ex_s - j \cdot d$.
\end{enumerate}

\end{lemma}

We exploit Lemma \ref{o:RepetativeLCP} to efficiently implement the following subroutine (details proof for the following can be found in Section \ref{A:complspace}.
\begin{lemma}\label{l:fastreplcp}
Given an arithmetic progression $A=(s,e,d)$ representing the indexes $\{s_j = i +j\cdot d|j \in [0\ldots |A|-1]\}$ that are contained within the same run with period $d$, and an index $s\in [1\ldots n]$. The values $lcp_j=LCP(s,s_j)$ can be evaluated and represented in $O(1)$ following $O(n)$ preprocess time on $S$.
\\\\
The representation consists of pairs $(I,L)$ such that $I=[a\ldots b]$ is a consecutive interval of $j$ values and $L$ is an integer such that one of the following holds:
\begin{enumerate}
    \item \label{type:const} $lcp_j=L$ for every $j\in I$ 
    \item \label{type:arit} $lcp_j = L -j\cdot d$ for every $j\in I$.
\end{enumerate}
Every pair is stored alongside with a bit indicating which one of the above holds for this pair.
\end{lemma}

In the process of evaluating the representation of $lcp_j$ for $A=(s,e,d)$ and $i$, at most one of the indexes in $A$ is called the aligned index. In the case in which $l_p < d$, the aligned index is $|A|-1$. In the case in which $l_p \ge d$, $j^*$ such that $Ex_s - j^* \cdot d$ is the aligned index, provided that it is an integer. We mark the pair representing the LCP value of the aligned index.

We employ Lemma \ref{l:fastreplcp} to obtain a representation of $l_1^j$ for every $j\in [0 \ldots |A|-1]$. After obtaining this representation, we are left with the task of applying a second $LCP$ query after the mismatch index for every $s_j$ (That will be the equivalent of finding $l_2$). Namely, for every $s_j$ we need to compute $l_2^j= LCP(i+l_1^j + 1,s_j + l_1^j + 1)$. More precisely , we need to count the number of $j$ values for which $l^j= l_1^j + l_2^j + 1 \ge m$.

For every pair $(I = [a \ldots b] , L)$, we wish to evaluate $l^j_2$ for $j \in I$ by employing Lemma \ref{l:fastreplcp} again. In order to do that, we first need to prove that the settings of Lemma \ref{l:fastreplcp} are satisfied in the second evaluation. We prove the following lemma in Section \ref{A:complspace}.

\begin{lemma}\label{l:conditionforfast}
For every pair $(I,L)$ in the output of Lemma \ref{l:fastreplcp} on $A=(s,e,d)$ and $i$ that is not corresponding to an aligned occurrence, one of the below holds for $j\in I$.
\begin{enumerate}
    \item \label{condition:1} $i+lcp_j+1$ is a fixed value and $s_j + lcp_j+1$ is an arithmetic progression of indexes within a run with period $d$ with difference $d$.
    \item \label{condition:2} $i+lcp_j+1$ is an arithmetic progression of indexes within a run with period $d$ with difference $d$ and $s_j + lcp_j+1$ is a fixed value 
\end{enumerate}
\end{lemma}

It follows from Lemma \ref{l:conditionforfast} that Lemma \ref{l:fastreplcp} can be applied to each of the pairs representing the non aligned indexes to evaluate a representation of $lcp^j_2$ for every non aligned index $j$ in $O(1)$ time. The aligned index, if exists, has its $l_2^{j}$ evaluated individually.

The above process outputs a set of (at most) 4 non-singular intervals for which the values of $l_2^j$ and $l_1^j$ are represented either as an arithmetic progression or as a fixed value. We can easily deduce the amount of occurrences $s_j$ with $l_1^j + l_2^j + 1 \ge m$ from this representation. We sum the amount of occurrences of $w_L$ with $l_1^j + l_2^j + 1 \ge m$ over all the arithmetic progressions of occurrences to obtain $|Lo(w)|$. A symmetric procedure can be constructed to evaluate $|Ro(w)|$ using occurrences of $w_R = w[\lceil \frac{m}{2} + 1 \rceil \ldots m]$. We can use the suffix tree to obtain $\#w$ (the number of occurrences of $w$ within $S$) for every $m$-length word in $S$ in $O(n)$ time.
The number of occurrences of $w$ with at most one mismatch is $|Ro(w)| + |Lo(w)| - \#w$. substracting $\#w$ is required to omit double counting. We do this process for every word of size $m$.

\textbf{Complexity.} For a word $w$ of size $m$, we process the arithmetic progressions of occurrences of $w_L$. For every arithmetic progression $A$, we evaluate a representation of $lcp^j_1$ and $lcp^j_2$ in constant time using Lemma \ref{l:fastreplcp}. We deduce the number of $l$-occurrences corresponding to the occurrence of $w_L$ represented by $A$ from the representation of $lcp^j_1$ and $lcp^j_2$ in constant time. We execute a symmetric procedure to deduce the number of $r$-occurrences of $w$ as well. There are $O(\frac{n}{m})$ arithmetic progressions in periodic occurrences representation of $w_L$, so counting the 1-ham occurrences of a single word takes $O(\frac{n}{m})$. We do this for every word of size $m$, so it adds up to $O(\frac{n^2}{m})$. There is an additional $O(n)$ preprocessing time prior to the iteration on the words to enable $LCP$ queries, suffix tree construction and access to the periodic occurrences representation. The overall complexity is $O(\frac{n^2}{m} + n) = O(\frac{n^2}{m})$.

\subsection{Reducing the complexity to $O(\frac{n^2}{m^2} + n)$}
For reducing the complexity by a factor of $m$, we present a technique for obtaining  $|Lo(w)|$ for a batch containing $O(m)$ words in $O(\frac{n}{m}+m)$ time. Consider the consecutive set of words with length $m$ starting in the indices $[i \ldots i + \frac{m}{4}]$. For every word $w^t = S[i + \frac{m}{4} - t \ldots i + \frac{m}{4} - t + m - 1]$ with $t \in [0 \ldots \frac{m}{4}]$ in this set, the left half of $w^t$ denoted as $w^t_L = S[i + \frac{m}{4} - t \ldots i + \frac{m}{4} - t + \lceil \frac{m}{2} \rceil]$ contains the word $w^i_L = S[i + \frac{m}{4} \ldots  i +\frac{m}{2} - 1]$. We use the occurrences of $w^i_L$ to evaluate $|Lo(w^t)|$ for every $t \in [0 \ldots \frac{m}{4}]$ as we did in the previous section. A symmetric process can be constructed for computing $|Ro(w)|$.

We start by finding the arithmetic progression representation of the occurrences of $w^i_L$. For simpler notation, we denote the starting and ending indices of $w^i_L$ as $w^i_L = S[s_i \ldots e_i]$. For every cluster $A$ with occurrences $\{s_j = s + d \cdot j| j \in [0 \ldots |A|-1] \}$, let $r_1^j = lcp( s_j, s_i)$ and $r_2^j =lcp (s_j + r_1^j + 1, s_i + r_1^j + 1)$. As in the previous section, we use Lemma \ref{l:fastreplcp} to obtain a compact representation of $r^j = r_1^j + r_2^j + 1$ for every index $s_j$ represented by $A$.

Using the following Lemma ,that can be proved similarly to Lemma \ref{l:fastreplcp}, we obtain a compact representation of $l^j = LCS(s_i - 1,s_j - 1) $ for every $j\in [0 \ldots |A|-1]$.
\begin{lemma}\label{l:fastreplcs}
Given an arithmetic progression $A=(s,e,d)$ representing the indexes $\{s_j = i +j\cdot d|j \in [0\ldots |A|-1]\}$ that are contained within the same run with period $d$, and an index $s\in [1\ldots n]$. The values $lcs_j=LCS(s,s_j)$ can be evaluated and represented in $O(1)$ following $O(n)$ preprocess time on $S$.
\\\\
The representation consists of pairs $(I,L)$ such that $I=[a\ldots b]$ is a consecutive interval of $j$ values and $L$ is an integer such that one of the following holds:
\begin{enumerate}
    \item \label{type2:const} $lcs_j=L$ for every $j\in I$ 
    \item \label{type2:arit} $lcs_j = L + j\cdot d$ for $j\in I$.
\end{enumerate}
\end{lemma}

After obtaining the values of $l_j$ and $r_j$, our next task is deducing for every $s_j$ represented by $A$, what are the values of $t$ for which $s_j$ is corresponding to an $l$-occurrence of $w^t$. The following observation is the key for doing so.
\begin{observation}\label{o:perContribution}
$s_j - t$ is an $l$-occurrence of $w^t$ iff $r^j \ge m - t$ and $l^j \ge t$.
\end{observation}

Observation \ref{o:perContribution} allows us to associate every occurrence $s_j$ with a continuous interval $I=[a\ldots b]$ such that $s_j$ is extendable to an $l$-occurrence of $w^t$ for and only for $t\in I$.

We initialize a data structure $D$ for maintaining $\frac{m}{4}$ counters $C_0,C_1,C_2 \ldots C_{\frac{m}{4}}$. $C_t$ counts $l$-occurrences of $w^t$. Initially, $C_t = 0$ for every $t \in [0\ldots \frac{k}{4}]$. We already know from observation \ref{o:perContribution} that for every $s_j$, the indexes $C_t$ with $t\in [m-r_j \ldots l_j]$ need to be increased by $1$. We call this type of updates, in which a consecutive interval of counters is increased by a constant value, an interval increment update. There are folklore techniques for applying this kind of updates to an array of counters efficiently.

Unfortunately, an efficient data structure for applying interval increment updates will not be sufficient for our cause, as we wish to process the effect of a \textbf{set} of occurrences on $D$. We therefore need to explore the structure of the set of updates $[m -r_j \ldots l_j]$ derived from the occurrences $s_j$ represented by a cluster $A$.

We present the following types of updates to be applied to an array of counters $D$.
\begin{definition}
An interval increment is represented by a triplet $(a,b,x)$. Applying $(a,b,x)$ to $D$ results in every counter $C_t$ with $t\in [a\ldots b]$ being increased by $x$.

An {\em increasing stairs update} is represented by a triplet $(a,b,p)$. The update requires applying the following modifications on $D$: 

For every $d\in 1 \ldots \lfloor \frac{b-a+1}{x} \rfloor$ Counters $C_i$ with $i \in [a + p \cdot (d-1) \ldots a+ p \cdot d - 1]$ are increased by $d$. The counters with $t\in [a + p \cdot \lfloor \frac{b-a+1}{x} \rfloor \ldots b]$ are increased by $\lfloor \frac{b-a+1}{x} \rfloor +1$

A {\em decreasing} stairs update is also represented by a triplet $(a,b,p)$. The update requires applying the following modifications on $D$: 

For every $d\in 1 \ldots \lfloor \frac{b-a+1}{x} \rfloor$ Counters $C_i$ with $i \in [b - p \cdot d +1 \ldots b - p \cdot (d - 1)]$ are increased by $d$. The counters with $t\in [a \ldots b - p \cdot \lfloor \frac{b-a+1}{x} \rfloor]$ are increased by $\lfloor \frac{b-a+1}{x} \rfloor +1$

We call the interval $[a \ldots b]$ the {\em span} of the stairs. We call the interval that is increased by $d$ the $d$th step of the update. $p$ is called the {\em width} of the stairs update. 

A {\em negative} stairs update (either increasing or decreasing) is a stairs update in which the counter in the $d$th step is decreased by $d$ rather than being increased by $d$.
\end{definition}

\begin{example}
Let $x = 10$ and an array $D = (0,0,0,0,0,0,0,0,0,0)$. An increasing stairs update $(3,8,2)$ on $D$ will result in the counters being set to $(0,0,1,1,2,2,3,3,4,0,0)$. Applying a decreasing stairs update $(1,5,2)$ on the updated counters in $D$ will result in the counters being set to $(3,2,3,2,3,2,3,3,4,0,0)$.
\end{example}

It turns out that a constant number of interval updates and stairs updates can be used to express the updates derived from the occurrences $s_j$ represented by a cluster $A$. In Section \ref{A:complspace}, we prove the following:
\begin{lemma}\label{l:representwithstairs}
Given a cluster $A$ of occurrences of $w^i_L$, the set of updates that need to be applied to $D$ in order to represent the $l$-occurrences corresponding to occurrences $s_j$ with $j\in [0\ldots |A|-1]$ can be represented by a constant number of stairs updates and interval increment updates. Given $A$ and the representation of $l_j$ and $r_j$, this set of stairs and interval increment updates can be retrieved in $O(1)$ time.

Over all the clusters representing occurrences of $w^i_L$, every stairs update $(a,b,p)$ in the representation has the same stairs width $p$ which is the period of $w^i_L$.
\end{lemma}

%xxxxxxx

Our algorithm runs as follows: Initialize a data structure $D$ for maintaining a set of $\frac{m}{4}$ counters. Find all the occurrences of $w^i_L$ in arithmetic progression representation. For every one of the $O(\frac{n}{m})$ arithmetic progressions, find the arithmetic progressions representing $r^j$ and $l^j$. Apply Lemma \ref{l:representwithstairs} to obtain an $O(1)$ size set of interval increment update and stairs update that represents the required modifications to be applied to $D$. The final ingredient for our algorithm is a data structure that enables the efficient application of these updates. In Section \ref{s:stairs}, we prove the following.
\begin{theorem}\label{t:stairs}
An array of $t$ counters can be maintained to support stairs updates in $O(1)$ time per update. Retrieving the values of all the counters in the array takes $O(t+u)$ time with $u$ being the amount of applied updates. The data structure works in the restricted settings in which every update $(a,b,p)$ has the same $p$ value. 
\end{theorem}

Note that the restriction on the queries hold in our case, since the step width is always $p$ the period of $w^i_L$ in all the stairs updates constructed in Lemma \ref{l:representwithstairs}. 

Every update corresponding to a set of occurrences of a certain type is applied to $D$ in $O(1)$ by employing the data structure of Theorem \ref{t:stairs} all the updates take $O(\frac{n}{m})$ by applying Theorem \ref{t:stairs}.

Note that we need a data structure for handling interval increment updates with the same complexities as the data structure of Theorem \ref{t:stairs}. The construction of such a data structure is quite simple and may be considered folklore. We therefore omit the implementation details of this data structure.

After applying the updates, we query our data structure for the values of all the counters. This process takes $O(m + \frac{n}{m})$ time. This is done for batches of $\frac{m}{4}$ consecutive indices. The indices of $S$ are partitioned to $4\frac{n}{m}$ such batches. We also preprocess the text for constant time $lcp$ and $lcs$ queries and construct the suffix tree. The total running time is $O(n + \frac{n}{m}(m + \frac{n}{m})) = O(n + \frac{n^2}{m^2})$.  Recall that we described a procedure for evaluating $|Lo(w)|$. A symmetric procedure can be constructed to evaluate $|Ro(w)|$.

Note that for $m\in \Omega(\sqrt{n})$, $O(\frac{n^2}{m^2} + n)$ is dominated by $O(n)$. The main result of this section immediately follows.
\begin{theorem}
For constant size alphabet and $m \in \Omega(\sqrt{n})$, the $1$-mappability problem can be solved in time $O(n)$.
\end{theorem}

\section*{Acknowledgements}
We warmly thank Tomasz Kociumaka for useful discussions.
%\section{Frequent Jumbled Subword with One Error}\label{s:jumbled}

\bibliographystyle{plainurl}
\bibliography{paper}

\newpage

\section{Appendix}

\subsection{Complementary Figures}
\begin{figure}[h]%\label{f:2}
    \centering
    %\scalebox{0.9}{\input{StairsDemonstration}}
      \includegraphics[width=120mm]{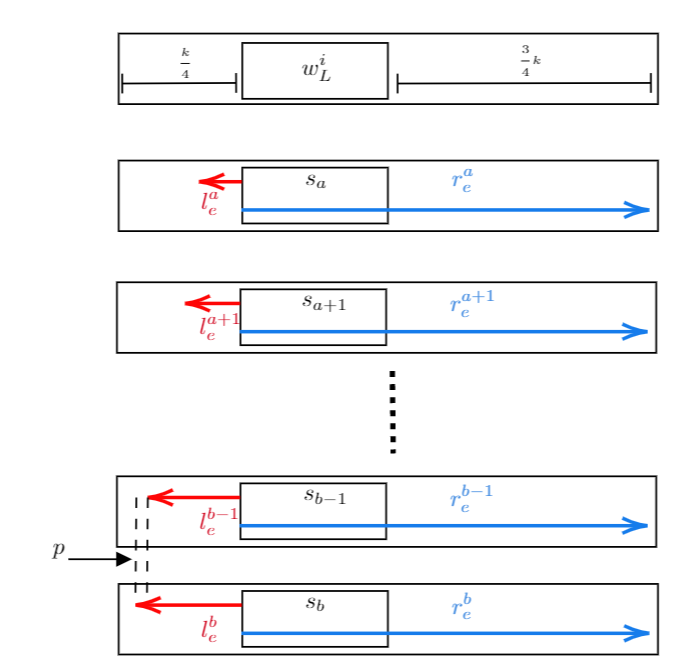}
    \caption{An illustration of the stairs update derived from a set of type 2 occurrences. Note that for every type 2 occurrence, the red arrow representing $l^j$ represents the interval of values of $t$ for which $C_t$ should be incremented due to an occurrence of $w_t$ in $s_j - t$. The $p$ indices that are only contained by the lowest step will be increased by $1$. The next $p$ indices are contained within two stairs and will be increased by 2. And so on.}
    \label{f:2}
\end{figure}
\eject
\begin{figure}[h]
    \centering
    %\scalebox{0.9}{\input{wleftdemonstrate.tex}}
      \includegraphics[width=120mm]{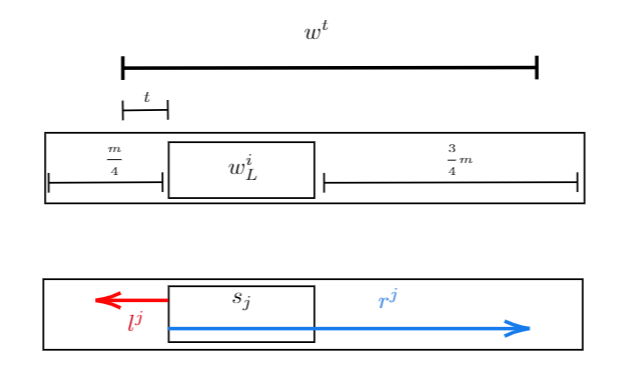}
    \caption{A demonstration of $w^i_L$ aligned with an occurrence $s_j$. Every $m$ sized word that fits within the interval spanned  by $l^j$ and $r^j$ (Red arrow and blue arrow, respectively) is an $l$-occurrence of a word $w_t$ that occurs in $s_j - t$}
    \label{f:1}
\end{figure}

\eject
\subsection{Complementary Proofs For Section \ref{s:lspace}}\label{A:complspace}
\textbf{Proof for Theorem \ref{t:apRep}:}
The existence of the representation specified by Theorem \ref{t:apRep} follows directly from the following facts:
\begin{fact}\label{f:aperoc}
An aperiodic string of length $m$ can have up to $O(\frac{n}{m})$ occurrences in a string $S$ of length $n$
\end{fact}
\begin{fact}\label{f:repocdist}
Let $w$ be a periodic word with period $p$ and length $k$. The distance between the starting points of two occurrences of $w$ in a string $S$ is either $p$ or greater than $\frac{m}{2}$.
\end{fact}

As for efficiently obtaining the periodic occurrences representation from the suffix tree, we present the following algorithm for preprocessing the suffix tree.
\\\\
{\bf Periodic Occurrences Representation Preprocess:}\\
\fbox{\begin{minipage}{12cm}

{\sf Input:} A suffix tree $ST$
\\\\
For every node $v\in ST$ with $|L(v)|=m$:
\begin{enumerate}
    \item {\sf Initialize} an empty list $L_v$ that is linked to $v$. 
    \item {\sf Initialize} a period $p_v= -1$
    \item {\sf Initialize} two auxiliary integers $pre_v = 0$ and $runstart_v=0$.
\end{enumerate}

{\sf Initialize} an array $A[1\ldots n - m + 1]$ with $A[i] = v$ such that $v$ is the node in $ST$ with $L(v)=S[i\ldots i+ m-1]$. 
\\\\
{\sf For every $i \in [1\ldots n]$:}
\begin{enumerate}
    \item Let $v=A[i]$
    \item If $pre_v=0$, set $pre_v=i$ and $runstart_v=i$.
    \item Otherwise:
    \begin{enumerate}
        \item If $i - pre_v > \frac{m}{2}$, add the pair $(runstart_v,pre_v)$ to $L_v$ and set $pre_v=i$ and $runstart_v=i$.
        \item Otherwise, set $i - pre_v= p_v$ and $pre_v = i$.
    \end{enumerate}
\end{enumerate}
{\sf For every $L_v$}:
\begin{enumerate}
    \item Add the pair $(runstart_v,pre_v)$ to $L_v$. 
    \item Replace every pair $(s,e)\in L_v$ with the tuple $(s,e,p_v)$.
    \end{enumerate}
\end{minipage}}

\begin{claim}\label{c:perrepcor}
$L_v$ contains the periodic occurrences representation of $L(v)$.
\end{claim}
\textbf{Proof:} For a periodic $L(v)$, the correctness of Claim \ref{c:perrepcor} directly follows from Fact \ref{f:aperoc} and the value of $p_v$ is irrelevant since every arithmetic progression will be a singleton. If $L(v)$ is periodic, every sequence of occurrences such that every occurrence starts $p=per(L(v))$ indexes to the right of the previous one will be represented as a single arithmetic progression. According to Fact \ref{f:repocdist}, the distance between the starting indexes of two such sequences of occurrences is at least $\frac{m}{2}$ , and therefore $|L_v| \in O(\frac{n}{m})$.
\QED

$A$ can be initialized in time $O(n)$ using the suffix tree. The rest of the algorithm is obviously linear. With that, the proof of Theorem \ref{t:apRep} is complete. \QED
\\\\

\textbf{Proof for Lemma \ref{o:RepetativeLCP}:} The correctness of the first case follows from the fact that every $s_j$ within the run has the same $d$ symbols to its right, possibly excluding the rightmost $s_j$.

As for the second case, note that the extension of the period from occurrence $s_j$ is $Ex_s - j\cdot d$. It holds that $S[i + x] = S[s_j + x]$ for every $d \le x < min(Ex_i,Ex_s - j\cdot d)$. This is due to the fact that for every such $x$ , $S[i + x - d] = S[i + x]$. And the first $d$ symbols to the right of $s_j$and $i$ are equal. If $Ex_i < Ex_s - j\cdot d$, The equality is broken in $S[i + Ex_i] \neq S[s_j + Ex_i]$ since $S[s_j + Ex_i] = S[s_j + Ex_i - d] = S[i + Ex_i- d] \neq S[i + Ex_i]$. Symmetrical arguments can be made for the case in which $Ex_i > Ex_s - j\cdot d$. \QED
\\\\

\textbf{Proof for Lemma \ref{l:fastreplcp}:}
We preprocess $S$ for constant time $LCP$ queries. Given $A$ and $s$, we evaluate $l_p=lcp(s,i)$ using an lcp query. We find the extension of the period $d$ to the right from $i$ and to the right from $s$ in constant time by querying $Ex_i =LCP(i, i+d)$ and $Ex_s = LCP(s ,s+d)$ respectively.
\\\\

If $l_p < d$, Observation \ref{o:RepetativeLCP} suggests that $lcp_j=l_p$ for $[1\ldots |A|-2]$. $lcp_{|A|-1}$ can be evaluated independently using an additional LCP query. Our representation consists of the pairs $([1\ldots |A|-2],l_p)$ and $(|A-1 \ldots |A|-1], l_{|A|-1})$. Both are pairs of type (\ref{type:const}).

In the case in which $l_p \ge d$, let $j^*$ be the number satisfying $Ex_i = Ex_s - j^* \cdot d$. The following fact is directly derived from Lemma \ref{o:RepetativeLCP}:
\begin{fact}
\begin{enumerate}
    \item $lcp_j =Ex_i $ for $j\in [0 \ldots min(\lceil j^*\rceil -1,|A|-1)]$
    \item $lcp_j= Ex_s - j\cdot d$ for $j\in [max(\lfloor j^* \rfloor + 1, 0) \ldots |A|-1]$
\end{enumerate}
\end{fact}
The above fact provides a representation for $lcp_j$ for every $j\neq j^*$. Specifically, the pair $([0 \ldots min(\lceil j^*\rceil -1,|A|-1)],Ex_i)$ of type (\ref{type:const}) and the pair $([max(\lfloor j^* \rfloor + 1, 0) \ldots |A|-1],Ex_s)$ of type (\ref{type:arit}). In the case in which $j^*$ is an integer, another pair of type (\ref{type:const}) with a singleton interval is required to represent $lcp_{j^*}$. $lcp_{j^*}$ can be independently evaluated using an LCP query.

The evaluation of $l_p$, $Ex_s$, $Ex_i$, $lcp_{|A|-1}$ and $lcp_{j^*}$ is done using a constant LCP query each and therefore consumes constant time. $j^*$ can be calculated from $Ex_s$, $Ex_i$ and $d$ using a constant number of basic arithmetic operations. The overall time for obtaining the representation of $lcp_j$ is constant. \QED
\\\\

\textbf{Proof of Lemma \ref{l:conditionforfast}:} in the case in which $l_p <d$, we have one pair $(I=[0\ldots |A|-2],l_p)$ that is corresponding to the non aligned occurrences. $i+l_p+1$ is a fixed value and $s_j + l_p+1$ is an arithmetic progression with difference $d$. Let $R$ be the run with period $d$ containing the indexes of $A$. Since $l_p<d$, and $I$ does not contain the rightmost index in the run , for every $j \in I$ $s_j$ has at least $d$ indexes to its right contained within $R$. Therefore, the index $s_j + l_p + 1 \le s_j + d$ is within $R$ for every $j\in I$ and condition (\ref{condition:1}) in the statement of the lemma holds.  

In the case in which $l_p \ge d$, we distinguish between the two pairs corresponding to the non-aligned indexes in the representation of $lcp_j$.

The indexes represented by the pair $(I=[0 \ldots min(\lceil j^*\rceil -1,|A|-1)], Ex_i)$ have $Ex_i< Ex_s -j \cdot d$. Since $lcp_j = Ex_i$ is a fixed value for $j\in I$, the sequence $i+ lcp_j + 1$ is fixed and $s_j + lcp_j + 1$ is an arithmetic progression with difference $d$. we also have $s_j + lcp_j + 1 = s_j + Ex_i + 1 \le Ex_s - j\cdot d = s + Ex_s$. Recall that $s+Ex_s$ is the right border of $R$, so $s \le s_j+lcp_j + 1 \le s_Ex_s$ suggests that $s_j + lcp_j +1$ is within $R$. We therefore proved that condition (\ref{condition:1}) holds in this case.

The indexes represented by the pair $(I=[max(\lfloor j^* \rfloor + 1, 0) \ldots |A|-1], Ex_s)$ have $Ex_i > Ex_s - j\cdot d$. Since $lcp_j = Ex_s - j\cdot d$ is an arithmetic progression with difference $-d$ for $j\in I$,  $i+lcp_j +1$ is an arithmetic progression with difference $-d$ and $s_j + lcp_j +1$ is a fixed value. Symmetric arguments to the ones in the previous case can be made to show that the indexes $i + lcp_j + 1$ are within the run with period $d$ containing $i$ and condition (\ref{condition:2}) holds.
\QED
\\\\
\textbf{Proof for Lemma \ref{l:representwithstairs}:}

We partition the occurrences $s_j$ into four distinct types:
\begin{enumerate}
    \item $s_j$ with $r^j \ge m$ and $l_e^j\ge \frac{m}{4}$. According to Observation \ref{o:perContribution},  $s_j - t$ is an $l$-occurrence of $w^t$ for every $t\in [0 \ldots \frac{m}{4}]$.
    \item $s_j$ with $r^j \ge m$ and $l^j < \frac{m}{4}$. According to Observation \ref{o:perContribution}, $s_j - t$ is an $l$-occurrence of $w^t$ for $t\in [0\ldots l^j]$.
    
    \item $s_j$ with $r^j < m$ and $l^j \ge \frac{m}{4}$. According to Observation \ref{o:perContribution}, $s_j - t$ is an $l$-occurrence of $w^t$ for $t\in [m - r^j\ldots \frac{m}{4}]$ in this case.
     
     \item $s_j$ with $r^j < m$ and $l^j \ge \frac{m}{4}$. According to Observation \ref{o:perContribution}, $s_j - t$ is an $l$-occurrence of $w^t$ for $t\in [m - r^j\ldots l^j]$ in this case.
\end{enumerate}
Fig.~\ref{f:1} demonstrates the fourth type listed above and can be used to understand the rest of the types. Recall that $r^j$ (resp. $l^j$) is partition into a constant number of intervals of values of $j$. For every such interval $I=[s\ldots e]$, an arithmetic progression represents the values of $r^j$ (resp. $l^j$) with $j \in I$. This representation can be easily processed in $O(1)$ time to obtain a partition $P$ of the values of $j$ into a constant number of intervals, such that every interval $I=[a\ldots b] \in P$ contains occurrences of exactly one of the types listed above.

We treat every type independently.

\textbf{Type 1:} An interval $I=[a \ldots b]$ of type 1 occurrences contributes $b - a + 1$ $l$-occurrences of $w^t$ for every $t \in [0 \ldots \frac{m}{4}]$. This is naturally represented by the interval increment update $(0, \frac{m}{4}, b-a +1)$

\textbf{Type 2:} Consider an interval $I=[a \ldots b]$ of type 2 occurrences. $s_j - t$ with $j\in I$ is an $l$-occurrence for every $w^t$ with $t\in[0 \ldots l^j]$. Recall $l^j$ is either an increasing arithmetic progression or a fixes value in $[a\ldots b]$. If it is a fixed value $l'$, every occurrence $s_j$ with $j\in I$ contributes an $l$-occurrence of $w_t$ for the same interval of $t$ values $ [0\ldots l']$. The overall contribution of all the occurrences in $I$ can be therefore represented with the interval increment update $(0,l',b-a+1)$.

The more complicated case is the case in which $l^j$ is an increasing arithmetic progression. Recall that the difference $p$ of this arithmetic progression is the period of $w^i_L$. The occurrence $s_b$ with the maximal LCP value $l^b$ contributes an $l$-occurrence of $w^t$ for $t\in [0 \ldots l^b]$. The occurrence $s_{b-1}$  contributes an $l$-occurrence for $w^t$ for $t\in [0 \ldots l^b - p]$ and so on. The effect of the entire progression on the counters $C_t$ can be described as follows: The counters $C_t$ for $t \in [l^b - p + 1 \ldots l_b]$ are increased by $1$, the counters with $t \in [l^b - 2p + 1 \ldots l^b - p]$ are increased by $2$ and so on. In general: the counters $C_t$ with $t \in [l^b- x\cdot p + 1 \ldots l^b - (x-1)\cdot p]$ are increased by $x$ for $x \in [1 \ldots b-a ]$ and the counters $C_t$ with $t\in [0 \ldots l^a]$ are increased by $b-a+1$. The modification of indexes in $[l^a + 1 \ldots l^b]$ can be equivalently described as an application of a decreasing stairs update $(l^a + 1,l^b,p)$. The modification of the indexes $[0\ldots l^a]$ can be described as an interval increment update $(0,l^a,b-a+1)$. See Fig.~\ref{f:2} for an illustration of the stairs update derived from type 2 occurrences.

\textbf{Type 3:} Having a symmetric structure to an interval of type 2 occurrences, the effect of an interval of type 3 occurrences on $D$ can be represented by a stairs update and an interval increment update as well.

\textbf{Type 4:} Consider a consecutive interval $I=[a\ldots b]$ of type 4 occurrences. Recall that, similarly to $l^j$, the arithmetic progression $r^j$ must be either decreasing or a fixed value. If both $l^j$ and $r^j$ are fixed in $I$, the counters $C_t$ with $t\in [m - r^b \ldots l^a]$ need to be increased by $a-b+1$ which can be represented with an interval increment update. If either $l^j$ or $r^j$ are fixed, and the other is an increasing or decreasing arithmetic progression, the required modification for $D$ can be represented with a stairs update and an interval increment update similarly to the representation of type 2 updates. %More precisely, if $r^j$ is a decreasing progression and $l^j$ is fixed, counters with $t\in [m - r^s .. m - r^s + p - 1]$ are increased by $1$. Counters with  $t\in [m - r^s + p ..m - r^s + 2p - 1]$ are increased by $2$ and so on until counters with $t \in [m - r^e .. l^e]$ are increased by $\lceil \frac{e-s+1}{p} \rceil$. 

If both $r^j$ and $l^j$ are arithmetic progressions, the updates to $C_t$ have a `sliding window' structure. Namely, Counters with $t \in [m - r^a \ldots l^a]$ are increased due to occurrence $s_a$. Counters with $t \in [m - r^a + p \ldots l_a + p]$ are increased due to occurrence $s_{a+1}$ and so on (Notice that these intervals may overlap). We proceed to show how to represent this kind of modification to the clusters using a constant number of stairs updates and interval increment updates.

For clearer presentation, assume that the required modification to be applied to the counters is given as a pair $(x,y)$ such that for every $j\in [0\ldots |I|-1]$ the interval $[x + j\cdot p \ldots y +j\cdot p]$ is increased by $1$. Every such interval $[x + j\cdot p \ldots y + j \cdot p]$ is called a window, with $x+j\cdot p$ being the start of the window and $y + j\cdot p$ being the end of the window. We represent the modification to the updates using two increasing stairs updates and one interval increment update.

The first increasing stairs update is $Starts= (x, x+ (|I|-1)\cdot p - 1 ,p)$. Note the the $d$-th step of $Starts$ starts in the same index as the start of the $d$-th window. The second update is a \textbf{negative} increasing stairs updates $Ends=(y+1, y+ (|I|-1)\cdot p,p)$. Note that the $d$-th step of $Ends$ starts one index to the right of the end of the $d$-th window. Finally, we have the interval increment update $Remainder = (x + (|I|-1)\cdot p , y +(|I|-1)\cdot p, |I|)$ which can be considered an extended last step for $Starts$. It is easy to see that all the updates only apply to the indexes affected by the sliding window. Furthermore, a counter $C_t$ is increased by $Starts$ (or by $Remainders$) by the number of starting indexes of windows that are not to the right of $t$. $C_t$ is decreased by $Ends$ by the number of windows with ending indexes strictly to the left of $t$. Overall, the counter $C_t$ is increased by the number of windows containing it. With this, we proved that $Starts$, $Ends$ and $Remainders$ are equivalent to the sliding window update given as $(x,y)$.

Every stairs or interval update we constructed in the above discussion can be easily obtained in $O(1)$ time from $I$ and from the representation of $l^j$ and $r^j$. The proof of the Lemma \ref{l:representwithstairs} is completed. \QED

\subsection{A data structure for implementing stairs updates}\label{s:stairs}

Let $D$ be a data structure for maintaining a set of $t$ counters denoted as $C_i,\ i \in [1 \ldots t]$.
\begin{definition}
A {\em stairs update} on $D$ is represented by a triplet $(a,b,p)$. The update requires applying the following modifications to $D$: 

Counters $C_i,\ i \in [a \ldots a+p - 1]$ are increased by 1 , Counters with $i \in [a+p \ldots a+2p - 1]$ are increased by 2 and so on until counters with $i\in [a+ \lfloor \frac{b-a+1}{p} \rfloor p \ldots b]$ are increased by $\lceil \frac{b-a+1}{p} \rceil$. 

We call the interval that is increased by $x$ by the stairs update the $x$'th step of the update. The above describes an {\em increasing} stairs update in which the value by which the indexes are increased is increasing from left to right within the interval $[a\ldots b]$. A {\em decreasing} stairs update is a stairs update in which the lowest step is adjacent to $b$. Specifically, counters $C_i$ with $i\in [b-p +1 \ldots b]$ are increased by $1$, counters with $[b-2p +1 \ldots b-p]$ are increased by $2$ and so on. We call the interval $[a \ldots b]$ the {\em span} of the stairs and $p$ the {\em width} of the stairs 
\end{definition}

\begin{example}
Let $x = 10$ and the indices of $D$ are $(0,0,0,0,0,0,0,0,0,0)$. An increasing stairs update $(3,8,2)$ on $D$ will result in the counters being set to $(0,0,1,1,2,2,3,3,4,0,0)$. Applying a decreasing stairs update $(1,5,2)$ on the updated counters in $D$ will result in the counters being set to $(3,2,3,2,3,2,3,3,4,0,0)$.
\end{example}

We proceed to introduce a data structure that enables the maintenance of $t$ indices under stairs updates. Our data structure works in the restricted settings in which all the updates have the same $p$ value. The time complexity for applying a stairs update is constant. The time for querying the values of all the indices is $O(t + u)$, with $u$ denoting the amount of updates.

We present a data structure for increasing stairs updates. A symmetrical construction can be made to achieve a data structure for decreasing stairs updates. Supporting both can be achieved by maintaining a data structure for increasing stairs updates alongside with a data structure for decreasing stairs updates and sum the corresponding counters.

Our data structure consists of two arrays $Start$ and $End$ of size $t$. Every cell in $Start$ contains an integer initialized to $0$. Every cell in $End$ contains an initially empty list of elements.

Upon stairs update $U = (a,b,p)$, We do the following: 
\begin{enumerate}
    \item increment $Start[a]$ by 1.
    \item Calculate the value $U$ adds to index $b$ : $U_{inc}=\lceil \frac{b-a+1}{p} \rceil$.
    \item Add the tuple $(U_{inc}, a \bmod p)$ to $End[b]$.
\end{enumerate}

Reporting the values of all the counters is done in a sequential manner:
Initialize an array of integers (initialized to 0) $Remainder$ of size $p$. Initialize $C=0$. Then do the following for $i=1 \ldots t$
\begin{enumerate}
    \item Increment $Reminders[i\bmod p]$ by $Start[i]$.
    \item Increment $C$ by $Reminders[i \bmod p]$ and report $C_i = C$.
    \item For every tuple $(U,r)$ in $End[i]$: Set $Remainders[r] = Remainders[r] - 1$ and $C = C - U$.
\end{enumerate}

\textbf{Correctness:} $Remainders[r]$ Contains the amount of `active' stairs updates that start in a position $a$ with $a = r \pmod p$. Therefore, The different between the previous $C_{i-1}$ and the current $C_i$ as a result of proceeding to a higher step within a stairs update is exactly $Remainders[i \pmod p]$. For the next value of $C_{i + 1}$, we subtract the contributions of the stairs update that ends in index $i$.

\textbf{Complexity} Every element in every array is inspected exactly once during the entire iteration, and it is handled in $O(1)$ time. The number of elements is bounded by $2u$, as for every applied update we add one element to $Start$ and one element to $End$. Additionally, we iterate from $i=1$ to $i = t$. So the overall complexity is $O(t + u)$.

The above concludes the proof of Theorem \ref{t:stairs}. 

\end{document}